# The efficacy of the sugar-free labels is reduced by the health-sweetness tradeoff

Ksenia Panidi[1*], Yaroslava Grebenschikova[1], Vasily Klucharev[1,2]

[1]Centre for Cognition and Decision Making, Institute for Cognitive Neuroscience, HSE University, Moscow, Russia

[2]Graduate School of Business, HSE University, Moscow, Russia

*Corresponding author

**Email**: kpanidi@hse.ru

**Postal address**: Centre for Cognition and Decision Making, Institute for Cognitive Neuroscience, HSE University, ul. Myasnitskaya 20, 101000, Moscow, Russian Federation

## Abstract

In the present study, we use an experimental setting to explore the effects of sugar-free labels on the willingness to pay for food products. In our experiment, participants placed bids for sugar-containing and analogous sugar-free products in a Becker-deGroot-Marschak auction to determine the willingness to pay. Additionally, they rated each product on the level of perceived healthiness, sweetness, tastiness and familiarity with the product. We then used structural equation modelling to estimate the direct, indirect and total effect of the label on the willingness to pay. The results suggest that sugar-free labels significantly increase the willingness to pay due to the perception of sugar-free products as healthier than sugar-containing ones. However, this positive effect is overridden by a significant decrease in perceived sweetness (and hence, tastiness) of products labelled as sugar-free compared to sugar-containing products. As in our sample, healthiness and tastiness are positively related, while healthiness and sweetness are related negatively, these results suggest that it is health-sweetness rather than health-tastiness tradeoff that decreases the efficiency of the sugar-free labelling in nudging consumers towards healthier options.

# Introduction

According to a recent World Health Organization report, noncommunicable diseases, among which are cardiovascular diseases and diabetes, are responsible for over 70% of avoidable premature deaths across the world (Chigom, 2020). Among the causes of these diseases is obesity, which particularly results from an overconsumption of ultra-processed foods that are rich in sugar (Moubarac et al., 2015; Popkin et al., 2012). Governments and commercial companies attempt to nudge consumers' dietary choices towards healthier options by placing the 'sugar-free' label on the food packaging containing natural or artificial sugar substitutes. However, little is known on whether this labeling is efficient in actually changing consumers' choices. In the present study, we report experimental results on the effect of the sugar-free label on the willingness to pay for the product. Additionally, we test the hypothesis that the effect of the sugar-free label on the willingness to pay is mediated by the health-taste tradeoff.

Consumers' dietary choices are, in part, directed by the information conveyed by nutritional labels on the food packaging (Campos et al., 2011), including labels informing consumers about the increased amount or absence of refined sugar (Scapin et al., 2021). However, the debate as to whether sugar-content related labelling is efficient in changing consumers' preferences is still ongoing. A recent systematic review found that sugar *warning* labels are effective in inducing healthier choices with respect to drinks high in sugar content (Gupta et al., 2021). Also, in a recent study there was a significant decrease in sugar-sweetened beverages choice due to warning labels on high-sugar drinks (Miller et al., 2022). Household purchases of soft drinks with high sugar contents dropped in Chile by 23.7% after the introduction of government sugar labelling regulations (Taillie et al., 2020). By contrast, in the field study by (Hoenink et al., 2021) no significant change was found in sales following the introduction of the colour labelling system with green labels for sugar-free and red for high sugar content. In another study, labels indicating the health-related properties of a product increased buying intention for that product only for a subgroup of participants (Bialkova et al., 2016).

Previous studies highlighted that labels may have ambiguous effects on consumption due to the health-taste trade-off that participants are subject to when evaluating labelled products. It was suggested that some consumers follow the so-called 'unhealthy=tasty' intuition when making judgments about the food's tastiness. In particular, they may believe that healthiness and pleasure of product consumption are inversely related (Raghunathan et al., 2006). They may also perceive a product as less tasty when health information is highlighted on the packaging (Grabenhorst et al., 2013). Additionally, for tasty products, labels highlighting hedonic properties have a more positive effect on evaluation compared to health labels (Fenko et al., 2016). Some studies also showed that following the 'unhealthy=tasty' intuition may lead to a higher body mass index by reducing the amount of vegetables in one's diet (Briers et al., 2020). By contrast, in other studies, a positive association between healthiness and tastiness has been reported (Haasova & Florack, 2019a, 2019b; Werle et al., 2013). It is also well established that consumers often base their healthiness and tastiness judgments on the cues related to the product packaging, such as brand name and package shape (Cavanagh & Forestell, 2013; van Ooijen et al., 2017). Non-directive product labels, i.e., labels that provide information about the nutrients, without prescribing any action based on this information, constitute an important cue for consumers as well (Hersey et al., 2013). However, in the presence of a health-taste tradeoff, labels such as 'sugar-free' may evoke opposing effects on consumers' judgments, increasing their perception of product healtiness, while decreasing its perceived tastiness. As a result, introducing such a label on the food packaging may not result in a greater willingness to pay for that product.

In the present study, we test the overall effect of the sugar-free label on the willingness to pay, as well as the components of this effect mediated by health, taste and sweetness considerations. In our experiment, participants placed bids in a Becker-deGroot-Marschak (BDM) auction (Becker et al., 1964), which is a well-established procedure used to derive a valuation of items (Asioli et al., 2021; Burchardi et al., 2021; Ginon et al., 2014; Hou et al., 2019). In this auction, a participant reports the maximum price that they are willing to pay for each product. Then, a random number is generated. If this random number is higher than the participant's bid, the participant does not receive the product and keeps the money. If this random number is lower than the participant's bid, the participant buys this product at a price equal to this number. The mechanism of this auction is such that it is optimal for a participant to reveal their true valuation of a product. We used real monetary rewards, and one trial was randomly selected at the end of the experiment to be played out for real. The stimuli were selected from among the products actually existing on the market.

Our main hypothesis was that the sugar-free label indicating absence of refined sugar in the product (but presence of natural sweeteners like stevia, honey, agave syrup, etc.) may increase the perceived healthiness of that product, thereby increasing its valuation for a consumer. Our experimental results support this hypothesis. However, and more importantly, we conclude that this positive effect of the label on product valuation is overridden by its negative effect on perceived product sweetness and tastiness, which results in an overall insignificant change in the willingness to pay for labelled products. In our dataset, healthiness and tastiness are positively related, while healthiness and sweetness are negatively related. Therefore, we show that the health-sweetness, rather than health-tastiness trade-off may hinder positive effects of non-directive sugar-related labels on product valuation.

## Materials and methods

*Participants*

Fifty participants (male=23, female=27), between 18 to 51 years (mean age=26.2, SD=6.9) completed the experiment. All participants met the standard criteria for participation in the behavioural experiment. The participants had normal or corrected to normal vision, did not have any neurological diseases, had no head injuries in the last 5 years and did not take psychotropic substances. All participants self-reported not having any psychological disorders related to food consumption (for example, an eating disorder), and did not have confirmed diabetes mellitus. Participants were not an expert in the field of dietetics (such as doctor, health coach, educator, etc.). Previous research shows that the deeper the knowledge a consumer possesses with respect to healthy eating habits – the more responsibly they behave when it comes to food choices, steering clear of questionable nutrients, like fast carbohydrates (Basil et al., 2009). Participants did not follow any diet or food restrictions during the last month and consumed sweet foods (e.g., sweets, chocolate, cookies) in everyday life. All subjects were asked not to eat anything 4 hours before the experiment, to unify the level of hunger (Epstein et al., 2003). The subjects received a reward of 300 monetary units (MU, ~13 USD, based on the BigMac index at the time of data collection), for participating in the experiment and a randomly selected product and an additional reward of up to 150 MU. The conversion rate for MUs to the local currency was 1:1. The subjects were recruited by email from the laboratory experiment participation database. The study was conducted at the Centre for Cognition & Decision Making (HSE university). All participants signed their informed consent before the beginning of the experiment. All procedures were approved by the ethics committee of the HSE University and were performed in accordance with relevant guidelines and regulations.

*Stimuli*

Since the regulation of the use of the sugar-free label is not legally restricted in the country of the experiment, the understanding of the 'sugar-free' labelling may differ among people. Therefore, to preselect the product categories, and to determine the most common intuitive meaning that people usually attach to the sugar-free label, an online survey on a separate group of participants was conducted prior to the main experiment. The online survey (N=100) aimed to determine the various product groups which people expect to contain the highest amount of sugar and/or sweeteners. In addition, the survey included several questions about the benefits and dangers of natural and artificial sweeteners.

The online survey was answered by 100 people (m=48, f=52) aged 18 to 60 years (mean age=33.4, SD=4.1) the popular online survey service. According to the results, the largest amount of refined sugar is thought to be contained in products such as chocolate, cakes and other confectionery groups (72% of participants) and cookies, crackers and other bakery groups (46% of participants). In the further study design, various products of these categories were used as stimuli. 51% of participants believe that natural sweeteners (fructose, honey, etc.) were healthier than refined sugar. 53% of participants believe that artificial sweeteners (xylitol, aspartame, etc.) were not healthier than refined sugar. 48% of participants believe that products labelled sugar-free do not contain refined sugar, but may contain natural sweeteners (honey, stevia). Based on these results, the group of products to be used in the experiment was limited to confectionery and bakery products. Sugar-free products were defined as those not containing refined sugar but potentially containing natural sweeteners (stevia, fructose, etc.).

The set of 60 products was then selected from the categories defined above. All products existed in the real market. Each of the products displayed cost up to 150 monetary units (MU). Of these 60 products, 30 items contained sugar and 30 items were sugar-free. Each sugar-containing product had a corresponding sugar-free product of a closely similar appearance when presented without packaging (which was the case in the experimental task).

All products were photographed without packaging with a Canon EOS 200D (resolution: 1920 × 1200 pixels) camera. The packaging was removed since previous studies have found that the presence of packaging, its colour, shape, and company label might influence consumers' behaviour (Okamoto & Dan, 2013, Campos et al., 2011). The photographs of each product were made from two different angles, and the presentation of the photographs was randomised across participants. The stimuli were presented on a white background (resolution: 800 × 600 pixels).

Sugar-free products were labelled with the sugar-free sign. To avoid the effect of colours on the perception of the stimuli (Visschers et al., 2010), the label represented a white circle with a black outline and the text 'SUGAR FREE' located inside the circle. The label appeared on the screen 1 second after the presentation of a product and stayed on the screen until the end of the stimulus presentation. The label was horizontally aligned to be in the centre and above the product picture. The position in the centre was chosen as the most noticeable for the attention of the participant according to previous studies (Graham & Jeffery, 2011, Visschers et al., 2010). The stimuli were presented using PsychoPy software (version 3.5). The example of stimuli is shown in Figure 1.

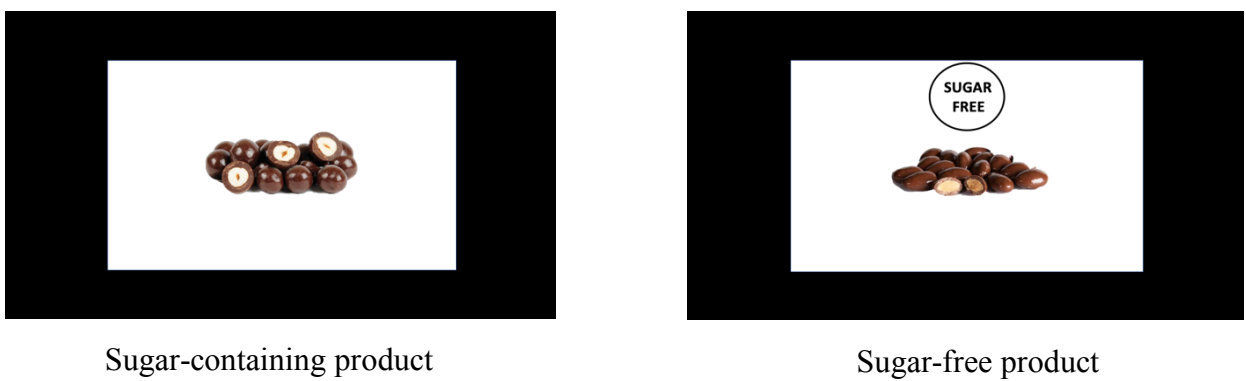

Sugar-containing product                                   Sugar-free product

**Figure 1.** Example of stimuli: sugar-containing product (left), analogous sugar-free product (right).

Experimental instructions explained the meaning of the sugar-free label as indicating that the product does not contain refined sugar but may contain natural sweeteners (such as honey, fructose, etc.), and that absence of the label indicated that the product contains refined sugar.

*Procedure*

The experiment consisted of two parts. In the first part, each participant went through the Becker–DeGroot–Marshak (BDM) auction to indicate their Willingness-To-Pay (WTP) for each product (Becker et al., 1964). In the second part, participants were again presented with the same products (in a different random order than in the first part) and had to indicate on a 5-point Likert scale, their familiarity with the product (1-not familiar at all, 5-very familiar), its perceived healthiness (1-very unhealthy, 5-very healthy), sweetness (1-not sweet at all, 5-very sweet) and tastiness (1-not tasty at all, 5-very tasty).
The order of questions was randomised for each product.

The two parts of the experiment were separated with a break of 10 minutes in between.

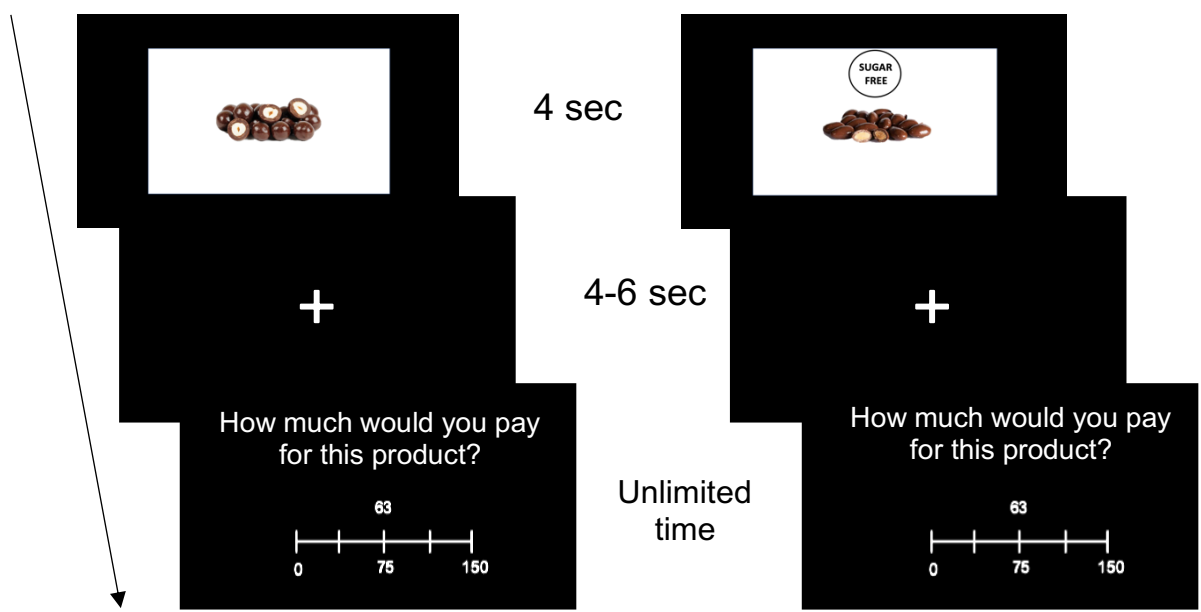

**Figure 2.** Trial structure for sugar-containing (left) and sugar-free (right) product.

*BDM auction*

In the BDM auction, each product was shown for 4 seconds, after which a white fixation cross appeared in the middle of the screen for 4-6 seconds (randomised across trials). After that, a participant was asked to set a bid for the product. The bid could be selected on the scale from 0 to 150 MU with a step of 1 MU. This scale was selected to ensure that the typical market price for all products would fit within this range. To select a bid, it was necessary to use the mouse cursor on the slider and press the "spacebar" to confirm the selection. Participants were informed that they should set a bid for the quantity of the product shown in the picture. There was no time limit for setting a bid. The trial structure is shown in Figure 2.

At the beginning of the experiment, 150 MU (in addition to the participation fee) were transferred to a participant's bank account to be used for the purchase of products. This was done to facilitate the perception of this endowment as money already owned by participants, which would promote more thoughtful decision-making during the experiment. Participants were also shown the box containing all the products, to make sure they understood that the experimental task was not hypothetical.

Participants were informed that the decisions they would make during the BDM auction would influence the additional reward they would get at the end of the experiment. The following conditions determined this additional reward. At the end of the experiment, one out of 60 products was randomly selected. Then a participant randomly pulled a capsule, with a number from 1 to 150 (step of 1) from the lottery urn. The number in the capsule was considered as the randomly selected price for the product. If the price offered by a participant for this product during the experiment was greater than or equal to the number from the capsule, participants bought the product for the price equal to the number from the capsule by transferring the corresponding amount back to the experimenter. The money that a participant did not spend on the purchase of the product remained in their account. If the price offered by participants for the product was less than the number from the capsule, participants did not receive the product, paid nothing, and kept 150 MU in full on their account. For example, assume the participant indicated a price of 83 MU for the selected product. The participant received the product if the number in the capsule was from 0 to 83. The participant did not receive the product if the number ball was from 84 to 150. The mechanism of the BDM auction was thoroughly explained to the participants in the instructions. It was also explained to them that it was in their best interest to offer a price at which they were really ready to buy the product.

*Pilot study*

As the main objective of the study was to determine the impact of the sugar-free label on the WTP for a product, we conducted a pilot study to test the hypothesis that the selected sugar-free and sugar-containing products do not differ in the WTP when no label is present. The rationale for this step comes from the fact that the WTP may be influenced by various factors, such as the expected taste, size, calorie content, etc. (Campos et al., 2011). Since the 30 sugar-free products were not completely identical to 30 corresponding sugar-containing products, the difference in the WTP, in the presence of labelling may then be incorrectly attributed to the label, while actually being generated by the differences in the appearance of the products. In order to avoid such distortions, a pilot experiment was conducted with the same set of products. In the pilot study, 30 products of the sugar-free category were shown without the sugar-free label, as well as the 30 sugar-containing products.

If the products in the sugar-free and sugar-containing groups are perceived similarly, when no information about sugar content is present, then we should observe no significant differences in the WTP between the two groups of products.

The procedure of the pilot experiment fully corresponded with the main experiment. The pilot experiment was attended by 12 people (m=5, f=7, mean age= 31), who fully met all the criteria for participants in the main experiment.

*Data analysis*

(1) Statistical tests

We first used a paired t-test to test the hypothesis that the average WTP is different between two product categories without controlling for any other product characteristics. The normality of the WTP distributions was confirmed with the Shapiro-Wilk test. Additionally, we used a Wilcoxon signed-rank test to test the hypothesis that products with and without label differ in the perceived familiarity, sweetness, healthiness and tastiness. The difference was considered significant when the p-value was below 0.05.

(2) Mediation analysis

To estimate the sugar-free label direct, indirect and total effects on the willingness to pay, we used the structural equation model (SEM). The structural equation modelling technique allows us to take into account the complex nature of interrelations between various characteristics of the stimuli (Bollen et al., 2013). The SEM technique allows us to set up a model where each variable can serve as a dependent variable in some equations, and as an explanatory variable in other equations. For example, in our study, perceived sweetness may directly affect willingness to pay, or it may indirectly increase it through increased perceived tastiness. Alternatively, it may decrease perceived healthiness and therefore negatively affect willingness to pay. The complete set of assumed relations between the product characteristics and the WTP is presented in **Figure 3**.

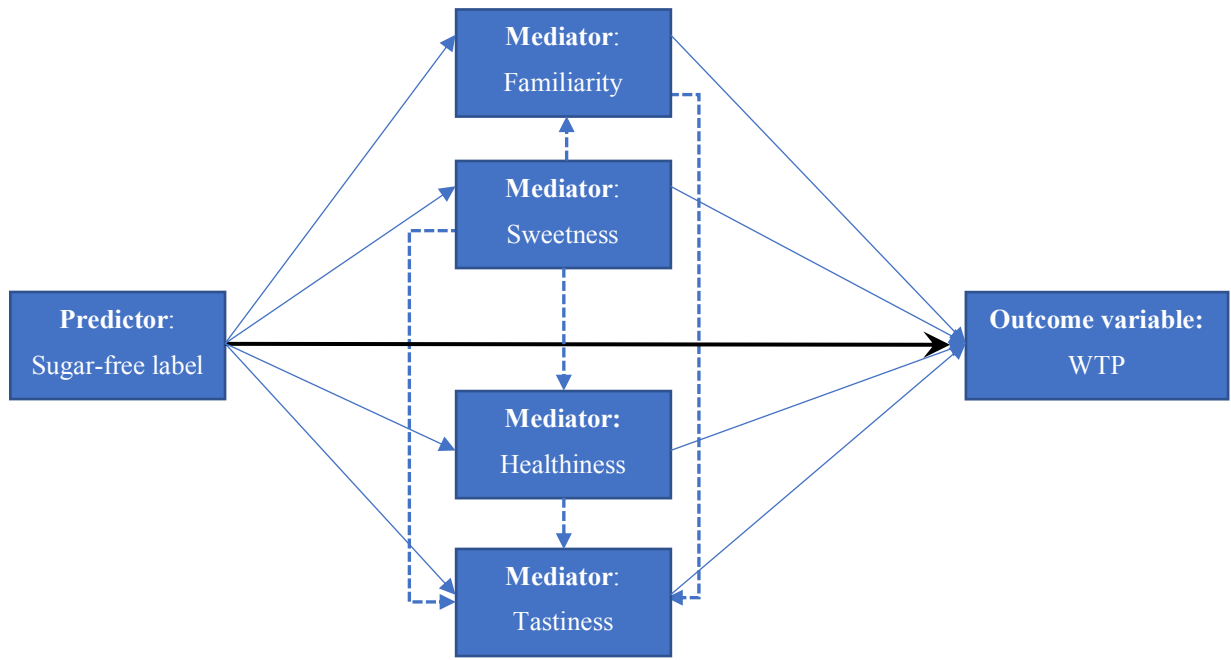

**Figure 3.** Path diagram describing the structural model for the effects of sugar-free labels on willingness to pay. The black arrow indicates the direct effect of the label on WTP. Solid blue arrows indicate the indirect effects of the label on WTP through each mediator variable. Dashed blue arrows indicate the relationship between mediator variables.

The structural equation model consisted of five equations. The dependent and explanatory variables for each equation are provided in **Table 1**. All equations included subject-level and product-level random effects. The product-level random effects were defined as random effects corresponding to each of the 30 pairs of sugar-free and analogous sugar-containing products.

| Regression number | Dependent variable | Explanatory variables | Control variables | Random effects |
|---|---|---|---|---|
| (1) | Familiarity | Sugar-free label<br>Sweetness | Gender, Age | Subject-level<br>Product-level |
| (2) | Sweetness | Sugar-free label | Gender, Age | Subject-level<br>Product-level |
| (3) | Healthiness | Sugar-free label<br>Familiarity<br>Sweetness | Gender, Age | Subject-level<br>Product-level |
| (4) | Tastiness | Sugar-free label<br>Familiarity<br>Sweetness<br>Healthiness | Gender, Age | Subject-level<br>Product-level |
| (5) | WTP | Sugar-free label<br>Familiarity<br>Sweetness<br>Healthiness<br>Tastiness | Gender, Age, Trial number | Subject-level<br>Product-level |

**Table 1.** Description of the structural model including interdependencies between product characteristics and the presence of the label.

We expected that familiarity with the product would be negatively affected by the label, because participants, in general, consume sugar-free products less frequently. For the same reason, we expected lower familiarity with products that are less sweet (the importance of this link for model fit was also confirmed by the test of directed separation (p-value<0.001)).

Perceived healthiness was expected to be positively affected by the label. Also, participants might be less familiar with healthier products, and perceive sweeter products as being less healthy due to the presence of refined sugar.

Perceived tastiness was expected to be negatively affected by the sugar-free label. Participants were expected to be more familiar with tastier products. Additionally, sweeter products were expected to be perceived as tastier, while the opposite was expected for healthier products.

Finally, the willingness to pay was expected to depend positively on the presence of the sugar-free label, as well as all four product characteristics.

To estimate this structural equation model, we used the piecewise SEM procedure (Zur et al., 2018). All calculations were performed in R v4.2.1 software (packages *piecewiseSEM* and *semEff* were used to estimate equation coefficients, as well as direct, indirect and total effects, and assess their significance). Fisher's C statistics was used to test for directed separation and evaluate the goodness-of-fit. Confidence intervals for the direct, indirect and total effects were obtained using bootstrapping with 1000 iterations. The effect was considered significant if zero was not included in the bootstrapped confidence interval.

# Results

*Pilot study results*

The results of the pilot study showed that there was no significant difference in WTP for sugar-free products and their analogous sugar-containing products (mean SF bid = 49.51, mean SC bid = 51.60, p-value 0.17).

The results also showed that sugar-containing products were perceived as significantly more familiar than their sugar-free analogues (mean for sugar-containing = 4.2, mean for sugar-free = 3.7, p-value = 0.007). In the other three characteristics, there was no significant difference. Both groups were almost identical in perceived sweetness (mean for sugar-containing = 3.7, mean for sugar-free =3.8, p-value = 0.82). Also, both product groups were the same in terms of perceived healthiness (mean for sugar-containing = 2.1, mean for sugar-free = 2, p-value = 0.84) and tastiness (mean for sugar-containing = 3.5, mean for sugar-free = 3.4, p-value = 0.63). The results are provided in **Figure 4**.

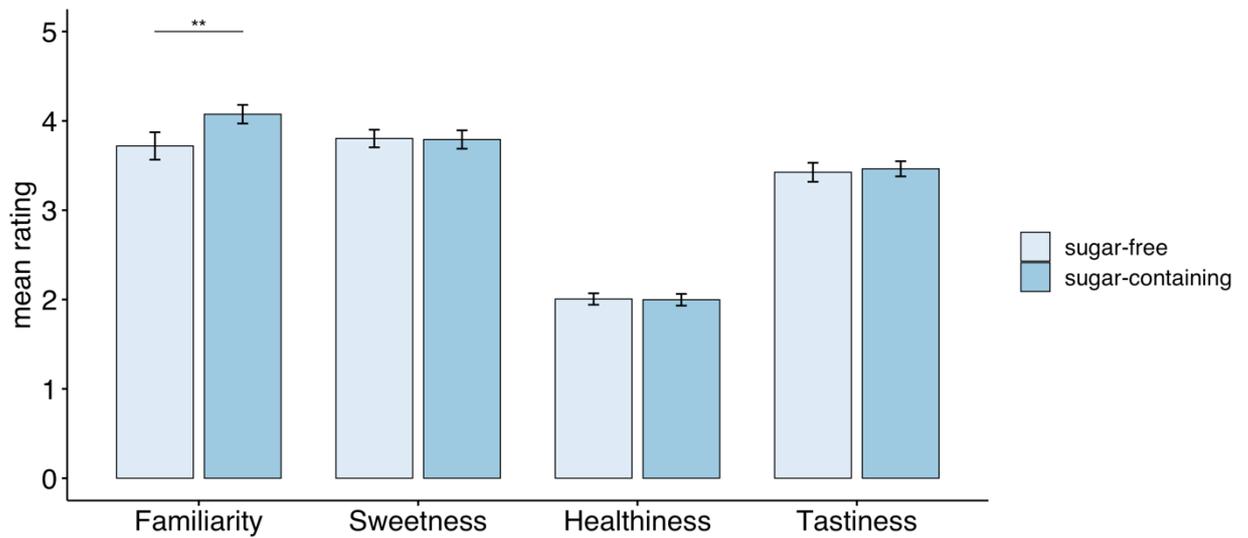

**Figure 4.** Pilot study results: average ratings of perceived familiarity, sweetness, healthiness and tastiness for sugar-containing and sugar-free products (in the pilot study, all products were presented without labels). The Wilcoxon signed-rank test was used to test the hypothesis that these characteristics are not different between the two product categories. Whiskers indicate the standard error of the mean. *** $p < 0.001$; ** $p < 0.01$; * $p < 0.05$.

*Main study: Regression analysis*

First, we analyzed the difference in WTP between sugar-free and sugar-containing products without taking into account any control variables. The average WTP was 41.95 MU for sugar-free products and 41.51 MU for sugar-containing ones. The paired t-test did not reveal any significant differences between the two product categories (p-value = 0.67).

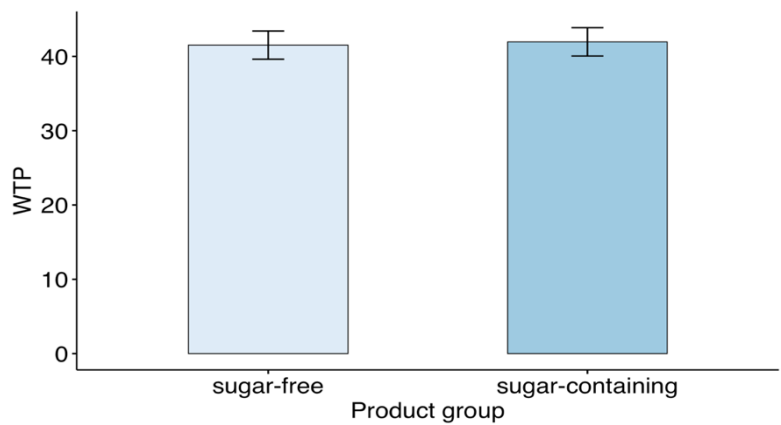

**Figure 5.** Main study results: average WTP for sugar-free products and their analogous sugar-containing products. There was no significant difference in WTP between the two product categories (p-value = 0.6709).

We observed significant differences in all four characteristics of products (see **Figure 5**). Sugar-containing products appeared significantly more familiar to participants (mean = 4.2; SD=1.3) compared to sugar-free products (mean=3.2; SD=1.1) (p-value<0.001). In the presence of the label, the products of the sugar-containing group were rated by the participants as significantly sweeter (mean = 3.9; SD=1.03) compared to sugar-free products (mean=3.3; SD=1.04) (p-value<0.001). Sugar-free products received a significantly higher score on healthiness (mean = 2.3; SD=0.8) compared to sugar-containing products (mean=1.9; SD=0.9) (p-value<0.001). The products of the sugar-containing group were rated by the participants as tastier (mean = 3.7; SD=1.2) compared to sugar-free products (mean=3.4; SD=1.2) (p-value=0.02).

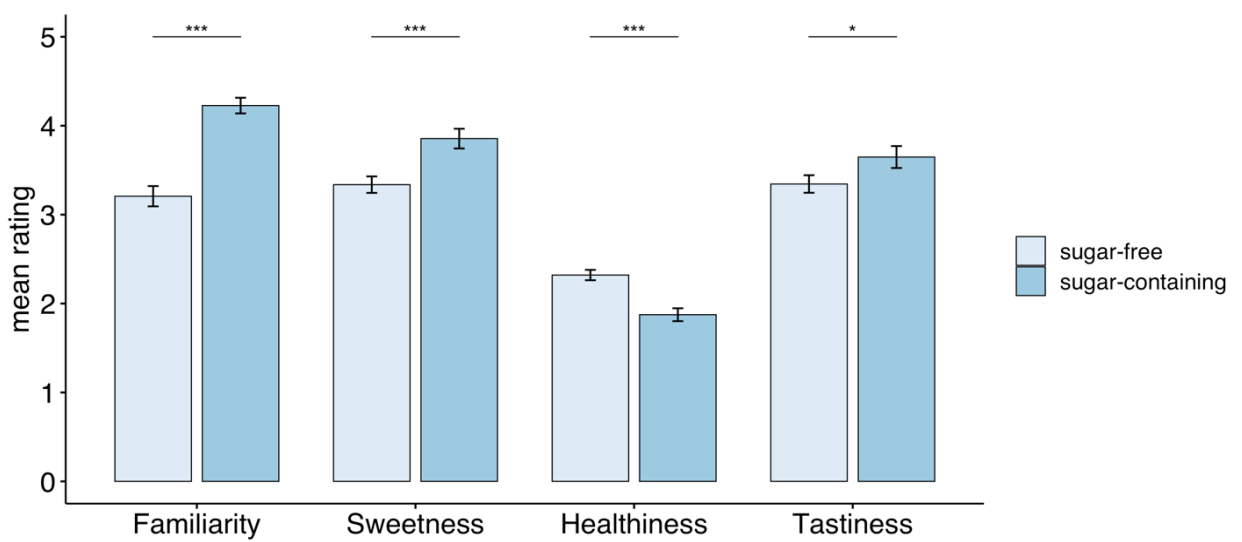

**Figure 6.** Main study results: average ratings of perceived familiarity, sweetness, healthiness and tastiness for sugar-containing and sugar-free products. The Wilcoxon signed-rank test was used to test the hypothesis that these characteristics are not different between the two product categories. Whiskers indicate standard error of the mean. *** p < 0.001; ** p < 0.01; * p < 0.05.

*Mediation analysis*

The piecewise SEM estimation results for each of the five model equations are presented in Table 2.

|  | Dependent variable: | | | | |
| --- | --- | --- | --- | --- | --- |
|  | (1) Familiarity | (2) Sweetness | (3) Healthiness | (4) Tastiness | (5) WTP |
| Constant | 3.161*** | 3.678*** | 3.081*** | 1.184*** | 1.445 |
|  | (0.256) | (0.241) | (0.271) | (0.255) | (9.658) |
| Sugar-free label (yes=1, no=0) | -0.877*** | -0.518*** | 0.353*** | -0.074 | 2.484** |
|  | (0.042) | (0.035) | (0.031) | (0.046) | (0.873) |
| Familiarity |  |  | 0.025* | 0.292*** | 1.274*** |
|  |  |  | (0.013) | (0.018) | (0.364) |
| Healthiness |  |  |  | 0.280*** | 4.352*** |
|  |  |  |  | (0.026) | (0.515) |
| Tastiness |  |  |  |  | 5.498*** |
|  |  |  |  |  | (0.350) |
| Sweetness | 0.275*** |  | -0.230*** | 0.110*** | 1.935*** |
|  | (0.021) |  | (0.015) | (0.022) | (0.433) |
| Trial number |  |  |  |  | -0.049* |
|  |  |  |  |  | (0.022) |
| Gender | 0.191 | 0.077 | 0.099 | 0.087 | 7.063 |
|  | (0.122) | (0.120) | (0.131) | (0.111) | (4.664) |
| Age | -0.003 | 0.006 | -0.018 | 0.009 | -0.094 |
|  | (0.009) | (0.009) | (0.009) | (0.008) | (0.332) |
| Observations | 3,000 | 3,000 | 3,000 | 3,000 | 3,000 |
| Log Likelihood | -4,643.879 | -4,199.497 | -3,546.321 | -4,628.289 | -13,487.770 |
| Akaike Inf. Crit. | 9,301.758 | 8,410.994 | 7,108.641 | 9,274.578 | 26,997.540 |
| Bayesian Inf. Crit. | 9,343.802 | 8,447.032 | 7,156.692 | 9,328.635 | 27,063.610 |

**Table 2.** Estimation results for structural equation model. The piecewise SEM procedure was used to estimate SEM equations. All regressions are mixed-effect linear models with subject-level and product-level random effects. *** $p < 0.001$; ** $p < 0.01$; * $p < 0.05$. Standard errors in parentheses.

Fisher's C statistic indicated that the model fits the data well ($F = 9,364$, p-value = 0.313) and there are no missing relationships between variables that should be included to explain the data. As expected, sugar-free products are less familiar to participants and are perceived as less sweet as well as more healthy.

We calculated the total, direct, and indirect effects of the sugar-free label on the willingness to pay with familiarity, sweetness, healthiness and tastiness as mediators. The results are presented in **Table 3**.

| Effect type | Variable | Effect estimate | Std. Err. | Lower CI | Upper CI | p-value |
|---|---|---|---|---|---|---|
| DIRECT | Sugar-free label | 2.357 | 1.018 | 0.542 | 4.627 | 0.021 |
| INDIRECT | Sugar-free label | -1.838 | 0.485 | -2.815 | -0.856 | < 0.001 |
| TOTAL | Sugar-free label | 0.520 | 1.055 | -1.366 | 2.852 | 0.622 |
| MEDIATORS | | | | | | |
| | Healthiness | 1.281 | 0.323 | 0.722 | 2.018 | < 0.001 |
| | Sweetness | -0.976 | 0.278 | -1.606 | -0.505 | < 0.001 |
| | Tastiness | -1.298 | 0.257 | -1.985 | -0.896 | < 0.001 |
| | Familiarity | -2.045 | 0.346 | -2.980 | -1.492 | < 0.001 |

**Table 3.** Direct, indirect and total effects of the sugar-free label on willingness to pay. The "Mediators" section presents the components of the total effect mediated by each of the four product characteristics. All effects are unstandardised (for standardised effects see Supplemental materials). Lower and Upper CIs correspond to the lower (2.5%) and upper (97.5%) confidence interval boundaries obtained by non-parametric bootstrapping. The effect was considered significant if CI does not contain zero.

The results show that the total effect of the sugar-free label on the willingness to pay is not significant (row 3 of Table 3, bootstrapped CI contains zero). However, this is due to the direct and indirect effects cancelling each other out (rows 1 and 2 in Table 3). The presence of the sugar-free label increases the willingness to pay by significantly increasing the perceived healthiness of the product, but decreases it via decreased sweetness, tastiness, and familiarity with the product. The significant positive direct effect of the label points at some other factors not included in the model but being important for explaining the overall label effect (see Discussion).

Interestingly, the regression estimation results in Table 5 suggest that there is a positive direct relationship between healthiness and tastiness. However, as sweetness is positively related to tastiness, and sugar-free labelled products are perceived as less sweet, we hypothesised that the negative effect of the label on tastiness is fully mediated by the sweetness component. The mediation analysis for the effects of the label on tastiness supports this hypothesis (**Table 4**).

| Effect type | Variable | Effect estimate | Std. Err. | Lower CI | Upper CI | p-value |
| --- | --- | --- | --- | --- | --- | --- |
| DIRECT | Sugar-free label | -0.034 | 0.037 | -0.114 | 0.035 | 0.358 |
| INDIRECT | Sugar-free label | -0.224 | 0.030 | -0.281 | -0.162 | < 0.001 |
| TOTAL | Sugar-free label | -0.258 | 0.036 | -0.333 | -0.189 | < 0.001 |
| MEDIATORS | | | | | | |
| | Healthiness | 0.095 | 0.018 | 0.064 | 0.132 | < 0.001 |
| | Sweetness | -0.072 | 0.024 | -0.127 | -0.030 | 0.003 |
| | Familiarity | -0.271 | 0.022 | -0.314 | -0.228 | < 0.001 |

**Table 4.** Direct, indirect and total effect of the sugar-free label on *Tastiness*. The "Mediators" section presents the components of the total effect mediated by each of the other product characteristics. All effects are unstandardised (for standardised effects see Supplemental materials). Lower and Upper CIs correspond to the lower (2.5%) and upper (97.5%) confidence interval boundaries obtained by non-parametric bootstrapping. The effect was considered significant if CI does not contain zero.

The mediation analysis for tastiness shows that the sugar-free label does not have a significant direct effect on perceived tastiness. The significant total effect stems from significant indirect effects mediated positively by perceived healthiness, and negatively by perceived sweetness and familiarity.

## Discussion

The present study tests the hypothesis that the sugar-free label increases the willingness to pay for a food product. In the present experiment, we used the Becker-deGroot-Marschak auction procedure to elicit the participants' willingness to pay in an incentive-compatible way. In this procedure, a participant states their bid for each product and then a random number is generated to indicate the product price. If a participant's valuation is higher than the random price, they get the product by paying the price, otherwise they do not get the product and keep the money. To increase the validity of the results, we used products actually existing on the real market. Additionally, the participants' choices in the experiment had real monetary consequences for them as they had an opportunity to either keep the endowment or to buy one of the products depending on their decisions.

The obtained results demonstrate that the sugar-free label increases willingness to pay via significantly increasing the perceived healthiness of the product, but at the same time the willingness to pay decreases because of lower perceived sweetness and familiarity with the sugar-free product. As these indirect effects act in opposing directions, the total effect of the label on willingness to pay turns out to be insignificant. Interestingly, we find that tastiness is directly positively associated with product healthiness, but negatively with sweetness. Hence, this evidence suggests a health-sweetness rather than health-tastiness tradeoff.

We also observe a direct effect of the label on WTP unexplained by the included mediators. We speculate that there might be several possible sources for this direct effect. First, since we deliberately used products existing on the real market, participants might be familiar with the typical prices of these products they could find in the supermarket. Therefore, in the process of selecting a bid they may act on the expectation that sugar-free products are usually more expensive than similar sugar-containing products. Although experimental instructions explained in detail how bidding would affect the probability of obtaining the product, the market price could still act as an unconscious anchor (Simonson & Drolet, 2004).

Second, the awareness about the negative consequences of excessive sugar consumption might have created an unconscious bias against the sugar-containing products which may not be fully captured by the indirect effect through perceived (un)healthiness. It has recently been shown that monetary losses may create both valuation bias (when an option is valued less when the possible loss is higher) and response bias (when an option is rejected, simply because it implies possible loss) (Sheng et al., 2020).A similar choice process, although in terms of health rather than monetary losses, may take place in our experimental task. For example, participants may demonstrate a response bias against sugar-containing products, unexplained by health value considerations.

Finally, the presence of the label as a distracting visual stimulus, itself, might have served as a source of bias in valuations of the sugar-free products. For example, some previous studies reported that inclusion of an irrelevant but salient stimulus into the stream of outcomes may lead to distorted valuations of these outcomes (Kunar et al., 2017). Various attentional processes may also affect the valuation of a product (Makarina et al., 2019). Further research is needed to clarify the nature of the direct effects of sugar-free labels on product valuations unexplained by perceived product characteristics.

Several limitations of the study should be mentioned.

We deliberately used colourless labels to explore the effect of information *directly* conveyed by the label without inducing any specific assessment of whether the absence of refined sugar is good or bad. For example, green coloured labels that are often found on food packaging may be perceived as indicating healthiness or a nudge towards favouring this product over sugar-containing ones (Emrich et al., 2017; Finkelstein et al., 2019; Kunz et al., 2020). By contrast, in most real-life situations consumers will see a coloured label which is likely to affect their opinions about product characteristics. Therefore, the overall effect of the label in this case may become significant and positive. The same concern relates to the fact that, in this study, we used food product photographs without any packaging to avoid any biases related to brand names and logos. However, in real life circumstances, these product properties may bias the health-taste or health-sweetness tradeoff in one or another direction, and, hence, shift the willingness to pay for the product. Finally, consumption habits in general, and awareness about the consequences of excessive sugar in particular, may differ across different cultures (Amerzadeh et al., 2022; Benitez et al., 2017). Therefore, further studies are needed to explore whether similar patterns will be observed in other western and eastern cultures.

The present study provides evidence that the sugar-free labelling may not be efficient in increasing the willingness to pay for sugar-free products due to the health-sweetness tradeoff faced by consumers. Although the label might have a positive direct effect on the participants' valuation, as well as indirect effect due to increased perceived healthiness of the product, this may be overridden by the opposing effects on how other product characteristics are perceived.

**Funding**: This article is an output of a research project implemented as part of the Basic Research Program at the National Research University Higher School of Economics (HSE University).

SUPPLEMENTAL MATERIAL

| Effect type | Variable | Effect estimate | Std. Err. | Lower CI | Upper CI | P-value |
|---|---|---|---|---|---|---|
| DIRECT | | | | | | |
| | Sugar-free label | 0.042 | 0.018 | 0.010 | 0.080 | 0.02 |
| INDIRECT | | | | | | |
| | Sugar-free label | -0.033 | 0.009 | -0.049 | -0.015 | < 0.001 |
| TOTAL | | | | | | |
| | Sugar-free label | 0.009 | 0.019 | -0.025 | 0.048 | 0.636 |
| MEDIATORS | | | | | | |
| | Healthiness | 0.023 | 0.006 | 0.013 | 0.035 | < 0.001 |
| | Sweetness | -0.017 | 0.005 | -0.029 | -0.009 | < 0.001 |
| | Tastiness | -0.023 | 0.005 | -0.035 | -0.016 | < 0.001 |
| | Familiarity | -0.036 | 0.006 | -0.050 | -0.027 | < 0.001 |

**Table S1.** Direct, indirect and total effect of the sugar-free label on willingness to pay (standardized). The "Mediators" section presents the components of the total effect mediated by each of the four product characteristics. Lower and Upper CIs correspond to the lower (2.5%) and upper (97.5%) confidence interval boundaries obtained by non-parametric bootstrapping.

| Effect type | Variable | Effect estimate | Std. Err. | Lower CI | Upper CI | P-value |
|---|---|---|---|---|---|---|
| DIRECT | | | | | | |
| | Sugar-free label | -0.014 | 0.014 | -0.044 | 0.014 | 0.317 |
| INDIRECT | | | | | | |
| | Sugar-free label | -0.089 | 0.012 | -0.112 | -0.063 | < 0.001 |
| TOTAL | | | | | | |
| | Sugar-free label | -0.103 | 0.014 | -0.131 | -0.075 | < 0.001 |
| MEDIATORS | | | | | | |
| | Healthiness | 0.038 | 0.007 | 0.026 | 0.052 | < 0.001 |
| | Sweetness | -0.029 | 0.010 | -0.051 | -0.012 | 0.004 |
| | Familiarity | -0.108 | 0.009 | -0.125 | -0.090 | < 0.001 |

**Table S2.** Direct, indirect and total effect of the sugar-free label on *Tastiness* (standardized). The "Mediators" section presents the components of the total effect mediated by each of the other product characteristics. Lower and Upper CIs correspond to the lower (2.5%) and upper (97.5%) confidence interval boundaries obtained by non-parametric bootstrapping.